\documentclass[aps,prd,onecolumn,showpacs,amsmath,amssymb,nofootinbib]{revtex4} \pdfoutput=1
 
\usepackage{amsfonts}
\usepackage{amsmath}
\usepackage{graphicx}
\usepackage{subfigure}
\usepackage{dcolumn}
\usepackage{bm}
\usepackage{booktabs}
\usepackage[utf8]{inputenc}
\usepackage{multirow}
\usepackage{graphicx,graphics,dcolumn,booktabs,bm}
\usepackage{longtable,lscape}
\usepackage{txfonts}
\usepackage{overpic}
\usepackage{amssymb}
\usepackage{indentfirst}
\usepackage{epsfig}
\usepackage{feynmf}   
\usepackage{epstopdf}   
\usepackage{slashed}  
\usepackage{color}
\usepackage[section]{placeins}

\usepackage[colorlinks, citecolor=blue,anchorcolor=red,menucolor=red, linkcolor=red,filecolor=red,runcolor=red,urlcolor=blue,frenchlinks=red]{hyperref}

\usepackage{ulem}

\newcommand{\dt}{\mathrm{d}t}
\renewcommand{\Im}{\mathrm{Im}}
\newcommand{\as}{\alpha_s}

\newcommand{\quarkfourd}{\langle m_s\overline{s}s\rangle}
\newcommand{\gluonfourd}{\langle\alpha G^2\rangle}
\newcommand{\mixed}{\langle g\overline{s} \sigma G s\rangle}

\newcommand{\angled}[1]{\langle #1\rangle}

\newcommand{\gev}{\ensuremath{\text{GeV}}}
\newcommand{\mev}{\ensuremath{\text{MeV}}}
\newcommand{\qcd}{\ensuremath{\text{QCD}}}
\newcommand{\msbar}{$\overline{\text{MS}}$}
\newcommand{\ie}{\textit{i.e.}}
\newcommand{\eg}{\textit{e.g.}}

\begin{document}
\title{Is the $Y(2175)$ a Strangeonium Hybrid Meson?}

\author{J.~Ho}
\affiliation{Department of Physics and Engineering Physics\\ University of Saskatchewan\\ 
          Saskatoon, SK, S7N 5E2, Canada}
\author{W.~Chen}
\affiliation{School of Physics\\ Sun Yat-Sen University\\ Guangzhou 510275, China}
\author{D.~Harnett}
\affiliation{Department of Physics\\ University of the Fraser Valley\\ 
          Abbotsford, BC, V2S 7M8, Canada}
\author{R.~Berg}
\affiliation{Department of Physics and Engineering Physics\\ University of Saskatchewan\\ 
          Saskatoon, SK, S7N 5E2, Canada}
\author{T.~G.~Steele}
\affiliation{Department of Physics and Engineering Physics\\ University of Saskatchewan\\ 
          Saskatoon, SK, S7N 5E2, Canada}

\begin{abstract}
\noindent QCD Gaussian sum-rules are used to explore the vector ($J^{PC}=1^{--}$) 
strangeonium hybrid interpretation of the $Y(2175)$.  
Using a two-resonance model consisting 
of the $Y(2175)$ and an additional resonance, we find that the relative
resonance strength 
of the $Y(2175)$ in the Gaussian sum-rules is less than 5\% 
that of a heavier $2.9\,\gev$ state.  
This small relative strength presents a challenge to a dominantly-hybrid interpretation
of the $Y(2175)$. 
\end{abstract}
\maketitle
\section{Introduction}\label{I}

The initial state radiation (ISR) process in $e^+e^-$ annihilation is a very useful technique to search for vector states (\ie, $J^{PC}=1^{--}$) in B-factories. 
In 2006, the BaBar Collaboration studied the cross sections for the ISR processes $e^+e^-\to K^+K^-\pi^+\pi^-$ and $e^+e^-\to K^+K^-\pi^0\pi^0$ up to $4.5\,\gev$,
aiming to confirm the existence of the $Y(4260)$ in the $\phi\pi\pi$ channels. 
Instead of observing the $Y(4260)$, however, they found a new resonance structure in the $\phi(1020)f_0(980)$ channel, which was named the $Y(2175)$~\cite{2006-Aubert-p91103-91103}. (It is also known as the $\phi(2170)$~\cite{2018-Tanabashi-p30001-30001}).
This resonance was later confirmed by BaBar~\cite{2007-Aubert-p12008-12008,2008-Aubert-p92002-92002,2012-Lees-p12008-12008}, BES~\cite{2008-Ablikim-p102003-102003}, and Belle~\cite{2009-Shen-p31101-31101} and recently by BESIII~\cite{2015-Ablikim-p52017-52017,2017-Ablikim-p-}. 
Its mass and decay width are $M=(2188\pm10)\,\mev$ and $\Gamma=(83\pm12)\,\mev$
and its quantum numbers are $I^GJ^{PC}=0^-1^{--}$~\cite{2018-Tanabashi-p30001-30001}.

To date, the nature of the $Y(2175)$ is still unknown. 
Based on strange quarkonium mass predictions using a relativized potential model,
only the $3\,{}^3S_1$ and $2\,{}^3D_1$ 
$s\bar{s}$ states are expected to have masses close to that 
of the $Y(2175)$~\cite{1985-Godfrey-p189-231}. 
However, both interpretations are disfavoured as the corresponding resonance width predictions 
are significantly larger than the width of the $Y(2175)$.  
The width of the $3\,{}^3S_1$ $s\bar s$ state 
was predicted to be $378\,\mev$ using the ${}^{3}P_0$ decay model~\cite{2003-Barnes-p54014-54014} 
whereas the width of the $2\,{}^3D_1$ $s\bar{s}$ state was predicted to be $167\,\mev$
in the ${}^3 P_0$ model and $212\,\mev$ in the flux tube breaking model~\cite{2007-Ding-p49-54}.
Another possible interpretation of the $Y(2175)$ is that of a strangeonium hybrid meson 
(\textit{i.e.,} $\bar{s}gs$).
Masses of vector strangeonium hybrid mesons have been computed using several methodologies
including the flux tube 
model~\cite{Merlin:1985mu,Isgur:1984bm,Merlin:1986tz,Barnes:1995hc}, 
lattice QCD~\cite{Dudek:2011bn}, 
and QCD Laplace sum-rules (LSRs)~\cite{Govaerts:1985fx}.
The flux tube model calculation of~\cite{Barnes:1995hc} found a 
vector strangeonium hybrid mass of 2.1--2.2$\,\gev$. 
The lattice QCD analysis of~\cite{Dudek:2011bn} found a vector strangeonium hybrid mass 
between $2.4\,\gev$ and $2.5\,\gev$ while
the LSRs calculation of~\cite{Govaerts:1985fx} found a heavier mass of $(2.9\pm0.3)\,\gev$.
Yet another possible interpretation of the $Y(2175)$ is that of a $ss\bar{s}\bar{s}$ tetraquark.
In~\cite{2018-Chen-p14011-14011}, the masses of vector $ss\bar s\bar s$ tetraquarks 
were investigated.  Two states 
were predicted with respective masses $(2.34\pm0.17)\,\gev$ and $(2.41\pm0.25)\,\gev$. 
Other LSRs analyses of $ss\bar{s}\bar{s}$ tetraquarks can be found 
in~\cite{2007-Wang-p106-116,2008-Chen-p34012-34012}.
Furthermore, 
the $Y(2175)$ has also been proposed as a molecular state of 
$\Lambda\bar\Lambda$~\cite{2010-Abud-p74018-74018,2013-Zhao-p54034-54034}.
In~\cite{2010-Abud-p74018-74018}, a chromomagnetic interaction Hamiltonian was 
used to predict a hexaquark of mass $2.184\,\gev$ that is strongly coupled to the
$\Lambda\bar{\Lambda}$ channel.
In~\cite{2013-Zhao-p54034-54034}, a one-boson-exchange potential model was used to predict
a $\Lambda\bar{\Lambda}$ mass between $2.149\,\gev$ and $2.181\,\gev$.
Also, the $Y(2175)$ has been interpreted as a dynamically generated resonance of 
$\phi f_0(980)$~\cite{2007-Napsuciale-p74012-74012,2008-MartinezTorres-p74031-74031,2009-Alvarez-Ruso-p54011-54011,2009-Gomez-Avila-p34018-34018}.

Decay modes and rates will be crucial to determining the nature of the $Y(2175)$.
In~\cite{2003-Barnes-p54014-54014}, it was predicted using the ${}^{3}P_0$ model
that the most important decay modes of the 
$3\,{}^3S_1$ $s\bar{s}$ meson would be $K^{*}K^{*}$, $KK^{*}(1410)$, and $KK_1(1270)$
whereas the $KK$ mode would be very weak.
In~\cite{2007-Ding-p49-54}, it was predicted using the ${}^{3}P_0$ model
that the most important decay modes of the
$2\,{}^3D_1$ $s\bar{s}$ meson would be $KK(1460)$, $KK^{*}(1410)$, $KK_1(1270)$, 
and $K^{*}K^{*}$.
While not dominant, the $KK$ decay mode was predicted to have a partial width of about 0.06.
In~\cite{2007-Ding-p390-400}, it was predicted using flux tube and constituent gluon models
that the most important decay modes of a vector strangeonium hybrid
would be $KK_1(1400)$, $KK_1(1270)$, $KK^{*}(1410)$, and $KK_2(1430)$, each
containing a $S$-wave meson plus a $P$-wave meson, due to the $S+P$ selection
rule~\cite{Isgur:1985vy,1999-Page-p34016-34016,2007-Ding-p390-400}. 
Also, it was noted that the $\phi f_0(980)$ mode could be significant.
Of particular interest are the $KK$, $K^{*}K^{*}$ and $KK(1460)$
modes which are predicted to be zero 
for a strangeonium hybrid interpretation (the usual rule that suppresses or even forbids
hybrid decays to pairs of S-wave mesons~\cite{Isgur:1985vy,Close:1994hc,Page:1996rj}).
For the $ss\bar s\bar s$ tetraquark interpretation, it has been suggested that the $\eta\phi$ channel should be one of the dominant decay modes due to the large phase space in the 
fall-apart mechanism~\cite{2007-Ding-p390-400}. 
However, in~\cite{2018-Ke-p-}, it was argued that the $\eta\phi$ decay mode would be greatly
suppressed and that the $\phi f_0(980)$, $h_1\eta$, and $h_1\eta'$ modes would be most important.
For the $\Lambda\bar{\Lambda}$ interpretation of the $Y(2175)$, the $KK$ decay mode
was predicted to dominate~\cite{2017-Dong-p74027-74027}.
At present, the data concerning decay modes and rates of the $Y(2175)$ is incomplete, making
it difficult to draw any definitive conclusions~\cite{2018-Tanabashi-p30001-30001}.

As they are both observed in ISR processes, 
the $Y(4260)$ and $Y(2175)$ states have the same quantum numbers, and 
are often considered as analogous states in the hidden-charm and hidden-strange sectors
respectively~\cite{Aubert:2005rm,2006-Aubert-p91103-91103,Ablikim:2016qzw}.
Perhaps determining the nature of one will shed light on the other.
Since the $Y(4260)$ has been identified as a good candidate for charmonium hybrid 
$\bar cgc$~\cite{Zhu:2005hp,Kou:2005gt,Palameta:2017ols}
or hidden-charm tetraquark state $qc\bar q\bar c$~\cite{2016-Chen-p1-121}, 
the $Y(2175)$ meson may also be interpreted as a hybrid or tetraquark candidate.

In this work, we use QCD Gaussian sum-rules (GSRs) methods to study the strangeonium hybrid possibility for $Y(2175)$.
In contrast to previous analyses of strangeonium hybrids using LSRs~\cite{Govaerts:1985fx}, 
the use of GSRs enables an exploration of the possibility of multiple states with hybrid components, allowing us to examine the scenario of a hybrid component of the $Y(2175)$.
We find little evidence in support of the $Y(2175)$ having a significant strangeonium
hybrid component.


\section{The Correlator and Gaussian Sum-Rules}\label{II}
We investigate vector strangeonium hybrids through the correlator
\begin{equation}
  \Pi(q^2) = \frac{i}{D-1}\left(\frac{q_{\mu}q_{\nu}}{q^2}-g_{\mu\nu}\right)\int d^{D}\!x\,e^{iq\cdot x} 
    \langle\Omega| \tau j_{\mu}(x) j_{\nu}(0) |\Omega\rangle
    \label{CorFnProj}
\end{equation}
where $D$ is spacetime dimension and where the current $j_{\mu}$ is given by
\begin{equation}\label{CurHyb}
  j_{\mu} = \frac{g_{s}}{2}\overline{s}\gamma^{\rho}\gamma_{5}\lambda^{a}
  \widetilde{G}^{a}_{\mu\rho}s.
\end{equation}
In~(\ref{CurHyb}),  
$s$ is a strange quark field and $\tilde{G}^a_{\mu\rho}$ is
the dual gluon field strength tensor,
\begin{equation}\label{dualFieldStrength}
  \widetilde{G}^{a}_{\mu\rho} = \frac{1}{2}\epsilon_{\mu\rho\omega\zeta}G^{a}_{\omega\zeta},
\end{equation}
defined in terms of the Levi-Civita symbol, $\epsilon_{\mu\rho\omega\zeta}$.

Between~\cite{Govaerts:1985fx} and~\cite{Ho:2018cat}, 
the quantity $\Pi(q^2)$ from~(\ref{CorFnProj}) 
has been computed to leading-order (LO) in $\as=\frac{g_s^2}{4\pi}$
within the operator product expansion (OPE).
In~\cite{Govaerts:1985fx}, the perturbative and dimension-four (\ie\ 4d)
quark and gluon condensate contributions were calculated.
In~\cite{Ho:2018cat}, the 5d mixed, 6d quark, and 6d gluon condensate contributions 
were calculated as well as $\mathcal{O}(m_s^2)$ corrections to 
perturbation theory where $m_s$ is the strange quark mass.
Denoting the result as $\Pi^{\qcd}(q^2)$ to emphasize that it is a QCD calculation, 
we have
\begin{equation}\label{CorOPE}
    \Pi^{\qcd}(q^2) = \left(\frac{\as}{\pi}
    \left(-\frac{q^6}{240\pi^2} + \frac{5m_s^2 q^4}{48\pi^2}  
    - \frac{4q^2}{9} \quarkfourd \right) + \frac{q^2}{36\pi} \gluonfourd 
    +  \frac{19\as m_s}{72\pi} \mixed \right)
    \log\left(\frac{-q^2}{\mu^2}\right)
\end{equation}
where
\begin{gather}
  \quarkfourd=\angled{m_s \overline{s}_i^{\alpha} s_i^{\alpha}}
    \label{condensateQuarkFour}\\
  \gluonfourd=\angled{\alpha_s G^a_{\mu\nu} G^a_{\mu\nu}}
    \label{condensateGluonFour}\\
  \mixed=\angled{g_s \overline{s}_i^{\alpha}\sigma^{\mu\nu}_{ij}
    \lambda^a_{\alpha\beta} G^a_{\mu\nu} s_j^{\beta}}
    \label{mixed}
\end{gather}
are respectively the 4d strange quark condensate, the 4d gluon condensate, 
and the 5d mixed strange quark condensate.
In~(\ref{condensateQuarkFour})--(\ref{mixed}), subscripts on strange quarks
are Dirac indices, superscripts are colour indices,
and $\sigma^{\mu\nu}=\frac{i}{4}[\gamma^{\mu},\,\gamma^{\nu}]$.
In computing~(\ref{CorOPE}), divergent integrals were handled 
through dimensional regularization in $D=4+2\epsilon$ dimensions at \msbar-renormalization 
scale $\mu$.
A dimensionally regularized $\gamma_5$ satisfying $\{\gamma_5,\gamma^{\mu}\}=0$ 
and $\gamma_5^2=1$ was used following the prescription of~\cite{Chanowitz:1979zu}.
Also, TARCER~\cite{Mertig:1998vk}, a Mathematica implementation of the recurrence 
relations of~\cite{Tarasov:1996br,Tarasov:1997kx}, was employed to reduce the set
of needed integral results to a small, well-known collection.
An irrelevant polynomial in $q^2$ has been omitted 
from~(\ref{CorOPE}) as it ultimately does not contribute to
the GSRs used in this article (see below).
Included in this omitted polynomial are the 6d quark and gluon condensate contributions,
both of which are constant for this channel as discussed in~\cite{Ho:2018cat}.

The quantity $\Pi(q^2)$ in~(\ref{CorFnProj}) is related to its imaginary part,
\ie, the hadronic spectral function, through a dispersion relation
\begin{equation}\label{dispersion}
  \Pi(Q^2) = \frac{Q^8}{\pi}\int_{t_0}^{\infty}\frac{\Im\Pi(t)}{t^4(t+Q^2)}\,\dt+\cdots
\end{equation}
at Euclidean momentum $Q^2\equiv -q^2>0$.
In~(\ref{dispersion}), $t_0$ is a hadron production threshold and $\cdots$ represents
subtraction constants, collectively a third degree polynomial in $Q^2$.
On the left-hand side of~(\ref{dispersion}), we identify $\Pi$ with 
$\Pi^{\qcd}$ of~(\ref{CorOPE}). 
On the right-hand side, we partition the hadronic spectral function
using a resonance-plus-continuum decomposition,
\begin{equation}\label{res_cont}
    \frac{1}{\pi}\Im\Pi(t) = \rho^{\text{had}}(t) + \theta(t-s_0)\frac{1}{\pi}\Im\Pi^{\qcd}(t),
\end{equation}
where $\rho^{\text{had}}(t)$ represents the resonance contribution to $\Im\Pi(t)$
and $\theta(t-s_0)$ is a Heaviside step function shifted to the 
continuum threshold parameter $s_0$.

In~(\ref{dispersion}), to eliminate subtraction constants 
as well as the aforementioned polynomials omitted from~(\ref{CorOPE})
and to enhance the resonance contribution relative to the continuum contribution
to the integral on the right-hand side, 
some transform is typically applied leading to some corresponding variant of QCD sum-rules.
Laplace sum-rules, for example, are a common choice 
(\eg, see~\cite{Shifman:1978bx,Shifman:1978by,Reinders:1984sr,Narison:2007spa}).
Here, we instead choose to work with (lowest-weight) GSRs
defined as~\cite{Bertlmann:1984ih}
\begin{equation}\label{definitionGSR}
  G(\hat{s},\,\tau) = \sqrt{\frac{\tau}{\pi}}
  \lim_{\stackrel{N,\Delta^2\rightarrow\infty}{\tau=\Delta^2/(4N)}}
  \frac{\big(-\Delta^2\big)^N}{\Gamma(N)}
  \bigg(\frac{d}{d\Delta^2}\bigg)^N
  \left\{\frac{\Pi(-\hat{s}-i\Delta)-\Pi(-\hat{s}+i\Delta)}%
  {i\Delta}\right\}.
\end{equation}
Discussions of how to evaluate definition~(\ref{definitionGSR}) for a correlator
such as~(\ref{CorOPE}) can be found in~\cite{Bertlmann:1984ih,Orlandini:2000nv,Harnett:2000fy}.
Substituting~(\ref{res_cont}) into~(\ref{dispersion}) and 
applying~(\ref{definitionGSR}), we find
\begin{align}
  G^{\text{QCD}}(\hat{s},\,\tau) &\equiv \frac{1}{\sqrt{4\pi\tau}}\int_0^{\infty}\!
  e^{-\frac{(\hat{s}-t)^2}{4\tau}} \frac{1}{\pi}\Im\Pi^{\text{QCD}}(t)\,\dt
  \label{unsubtracted_GSR_LHS}\\
  \implies G^{\text{QCD}}(\hat{s},\,\tau)
  &= \frac{1}{\sqrt{4\pi\tau}}\int_{t_0}^{\infty}\!
  e^{-\frac{(\hat{s}-t)^2}{4\tau}} \rho^{\text{had}}(t)\,\dt
  + \frac{1}{\sqrt{4\pi\tau}}\int_{s_0}^{\infty}\!
  e^{-\frac{(\hat{s}-t)^2}{4\tau}} \frac{1}{\pi}\Im\Pi^{\text{QCD}}(t)\,\dt.
  \label{unsubtracted_GSR_RHS}
\end{align}
Subtracting the continuum contribution,
\begin{equation}\label{continuum}
  \frac{1}{\sqrt{4\pi\tau}}\int_{s_0}^{\infty}\!
  e^{-\frac{(\hat{s}-t)^2}{4\tau}} \frac{1}{\pi}\Im\Pi^{\text{QCD}}(t)\,\dt,
\end{equation}
from~(\ref{unsubtracted_GSR_LHS}) and~(\ref{unsubtracted_GSR_RHS})
leads to subtracted GSRs
\begin{align}
  G^{\text{QCD}}(\hat{s},\,\tau,\,s_0) &\equiv \frac{1}{\sqrt{4\pi\tau}}\int_0^{s_0}\!
  e^{-\frac{(\hat{s}-t)^2}{4\tau}} \frac{1}{\pi}\Im\Pi^{\text{QCD}}(t)\,\dt
  \label{subtracted_GSR_LHS}\\
  \implies G^{\text{QCD}}(\hat{s},\,\tau,\,s_0)
  &= \frac{1}{\sqrt{4\pi\tau}}\int_{t_0}^{\infty}\!
  e^{-\frac{(\hat{s}-t)^2}{4\tau}} \rho^{\text{had}}(t)\,\dt.
  \label{subtracted_GSR_RHS}
\end{align}
Finally, calculating $\Im\Pi^{\qcd}(t)$ from~(\ref{CorOPE}) and substituting the
result into the right-hand side of~(\ref{subtracted_GSR_LHS}), we find
\begin{equation}\label{GSRExplicit}
  G^\text{QCD}(\hat{s},\,\tau,\,s_0) \equiv \frac{1}{\sqrt{4\pi\tau}}\int_0^{s_0}\!
  e^{-\frac{(\hat{s}-t)^2}{4\tau}}
  \left(\frac{\as}{\pi}
  \left(-\frac{t^3}{240\pi^2} + \frac{5m_s^2 t^2}{48\pi^2}  
  - \frac{4t}{9} \quarkfourd \right) + \frac{t}{36\pi} \gluonfourd 
  +  \frac{19\as m_s}{72\pi}\mixed\right)\,\dt.
\end{equation}
Note that the definite integral in~(\ref{GSRExplicit}) can be evaluated in terms 
of error functions.
The kernel of the subtracted GSRs is a Gaussian of width $\sqrt{2\tau}$ centred at $\hat{s}$.
As discussed in~\cite{Orlandini:2000nv,Harnett:2000fy,Harnett:2008cw,Ho:2018cat}, 
GSRs are particularly well-suited to the study of multi-resonance hadron models as, 
by varying $\hat{s}$, excited and ground state resonances can be probed with 
similar sensitivity.

Renormalization-group (RG) improvement of~(\ref{GSRExplicit}) amounts to replacing
$\alpha_s$ and $m_s$ by running quantities at the scale 
$\mu^2=\sqrt{\tau}$ (\eg, \cite{Bertlmann:1984ih,Narison:1981ts}).
The one-loop, \msbar\ running coupling at $n_f=4$ active quark flavours is
\begin{equation}\label{running_coupling}
\alpha_{s}(\mu) = \frac{\alpha_{s}\left(M_{\tau}\right)}%
{1+\frac{25}{12\pi}\as(M_{\tau})\log\left(\frac{\mu^2}{M_{\tau}^2}\right)}
\end{equation}
where \cite{2018-Tanabashi-p30001-30001}
\begin{gather}
  M_\tau = 1.77686 \pm0.00012\,\gev\\
  \alpha_{s}\left(M_{\tau}\right) = 0.325\pm 0.015.
\end{gather}
Since the previous analysis of strangeonium hybrid mesons using QCD sum-rules \cite{Govaerts:1985fx}, the condensate parameters and quark masses are now known more precisely. In addition to the inclusion of higher-dimensional condensates terms in \eqref{CorOPE}, we update the values and uncertainties in the QCD parameters used in \cite{Govaerts:1985fx}. The running strange quark mass is 
\begin{equation}\label{running_mass}
  m_s(\mu) = m_s(2~\gev)\left(\frac{\as(\mu)}{\as(2~\gev)}\right)^{\frac{12}{25}}
\end{equation}
where 
\cite{2018-Tanabashi-p30001-30001} 
\begin{equation}
   m_s(2~\gev)=96^{+8}_{-4}\ \mev.
\end{equation}

The value of the RG-invariant 4d strange quark condensate is known from PCAC,
\begin{equation}
  \quarkfourd = -\frac{1}{2} f_K^2 m_K^2,
\end{equation}
where \cite{2018-Tanabashi-p30001-30001, Rosner:2015wva}
\begin{gather}
  m_K = (493.677\pm0.016)\ \mev\\
  f_K = (110.0 \pm 4.2)\ \mev
\end{gather}
For the 4d gluon condensate, we use the value from~\cite{Narison:2011rn},
\begin{equation}
    \gluonfourd = (0.075 \pm 0.020)\ \gev.
\end{equation}
For the 5d mixed strange quark condensate, we use the estimate from~\cite{Beneke:1992ba,Belyaev:1982sa},
\begin{equation}
    \frac{m_s\mixed}{\quarkfourd} \equiv M_0^2 = (0.8 \pm 0.1)\ \gev^2.
\end{equation}

Integrating~(\ref{subtracted_GSR_RHS}) with respect to $\hat{s}$ gives
\begin{equation}\label{fesr_constraint}
  \int_{-\infty}^{\infty} G^{\qcd}(\hat{s},\,\tau,\,s_0)\,\mathrm{d}\hat{s}
  = \int_{t_0}^{\infty} \rho^{\text{had}}(t)\,\mathrm{d}t.
\end{equation}
The quantity on the LHS of~(\ref{fesr_constraint}) is the lowest-weight finite energy sum-rule (FESR), and, as shown in~\cite{Bertlmann:1984ih}, the spectral function
decomposition~(\ref{res_cont}) only reproduces the QCD prediction at high
energy scales if $s_0$ is constrained by~(\ref{fesr_constraint}).
To isolate the information in the GSRs that is independent of the 
FESR constraint~(\ref{fesr_constraint}), we define normalized 
GSRs (NGSRs)~\cite{Orlandini:2000nv},
\begin{equation}
    N^\text{QCD}(\hat{s},\,\tau,\,s_0) \equiv \frac{G^\text{QCD}(\hat{s},\,\tau,\,s_0)}%
    {\int_{-\infty}^{\infty} G^\text{QCD}(\hat{s},\,\tau,\,s_0)\,\mathrm{d}\hat{s}}
    \label{NGSR}
\end{equation}
which, from~(\ref{subtracted_GSR_RHS}) and~(\ref{fesr_constraint}), 
implies that
\begin{equation}\label{subtracted_NGSR}
    N^\text{QCD}(\hat{s},\,\tau,\,s_0) = 
    \frac{\frac{1}{\sqrt{4\pi\tau}}\int_{t_0}^{\infty}\!
    e^{-\frac{(\hat{s}-t)^2}{4\tau}}\rho^{\text{had}}(t)\,\dt}%
    {\int_{t_0}^{\infty}\rho^{\text{had}}(t)\,\mathrm{d}t}.
\end{equation}

\section{Analysis Methodology and Results}\label{IV}
Previous work using LSRs used a single-narrow resonance model and resulted 
in a strangeonium hybrid mass prediction significantly heavier than the $Y(2175)$~\cite{Govaerts:1985fx}.
Compared with that analysis, we include 5d and 6d condensate terms in the OPE
and use updated QCD parameter values.  Also, as outlined above, Gaussian sum-rules have the ability to probe multiple states in the spectral function.  We can therefore update and extend the analysis of Ref.~\cite{Govaerts:1985fx} and test the hypothesis of a $Y(2175)$ hybrid component by using a double-narrow resonance model for the hadronic spectral function
\begin{equation}
    \rho^{\text{had}}(t)=f_1^2\delta\left(t-m_1^2 \right)+
    f_2^2\delta\left(t-m_2^2 \right)\,.
\end{equation}
This double narrow-resonance model   in 
 \eqref{subtracted_GSR_RHS} provides the hadronic contribution,
 \textit{i.e.,} the right-hand side, to the NGSRs~(\ref{subtracted_NGSR}),
\begin{equation}\label{gsrDNR}
  N^\text{had}\left( \hat{s},\,\tau \right)= \frac{1}{\sqrt{4\pi \tau}}\Bigg( r e^{-\frac{(\hat{s} - m_{1}^{2})^2}{4\tau}}
  +  (1-r) e^{-\frac{(\hat{s} - m_{2}^{2})^2}{4\tau}} \Bigg),
\end{equation}
where the normalized couplings are defined as
\begin{equation}\label{rDefn}
 r= \frac{f_{1}^{2}}{f_{1}^{2}+f_{2}^{2}}\,,
 1-r = \frac{f_{2}^{2}}{f_{1}^{2}+f_{2}^{2}}\,, 0\le r\le 1.
\end{equation}
We fix one of our modelled resonances ($m_1$) using the experimental value given in 
Refs.~\cite{2018-Tanabashi-p30001-30001,ChenHX2018},
\begin{equation}
    m_1 = m_{Y(2175)} = 2.188\ \gev,
\end{equation}
and the additional resonance ($m_2$) provides the necessary degrees of freedom in the model 
for the possibility that the $Y(2175)$ decouples 
(\textit{i.e.,} that $m_1$ has normalized coupling $r\approx 0$).

\begin{figure}
    \centering
    \includegraphics[scale=0.75]{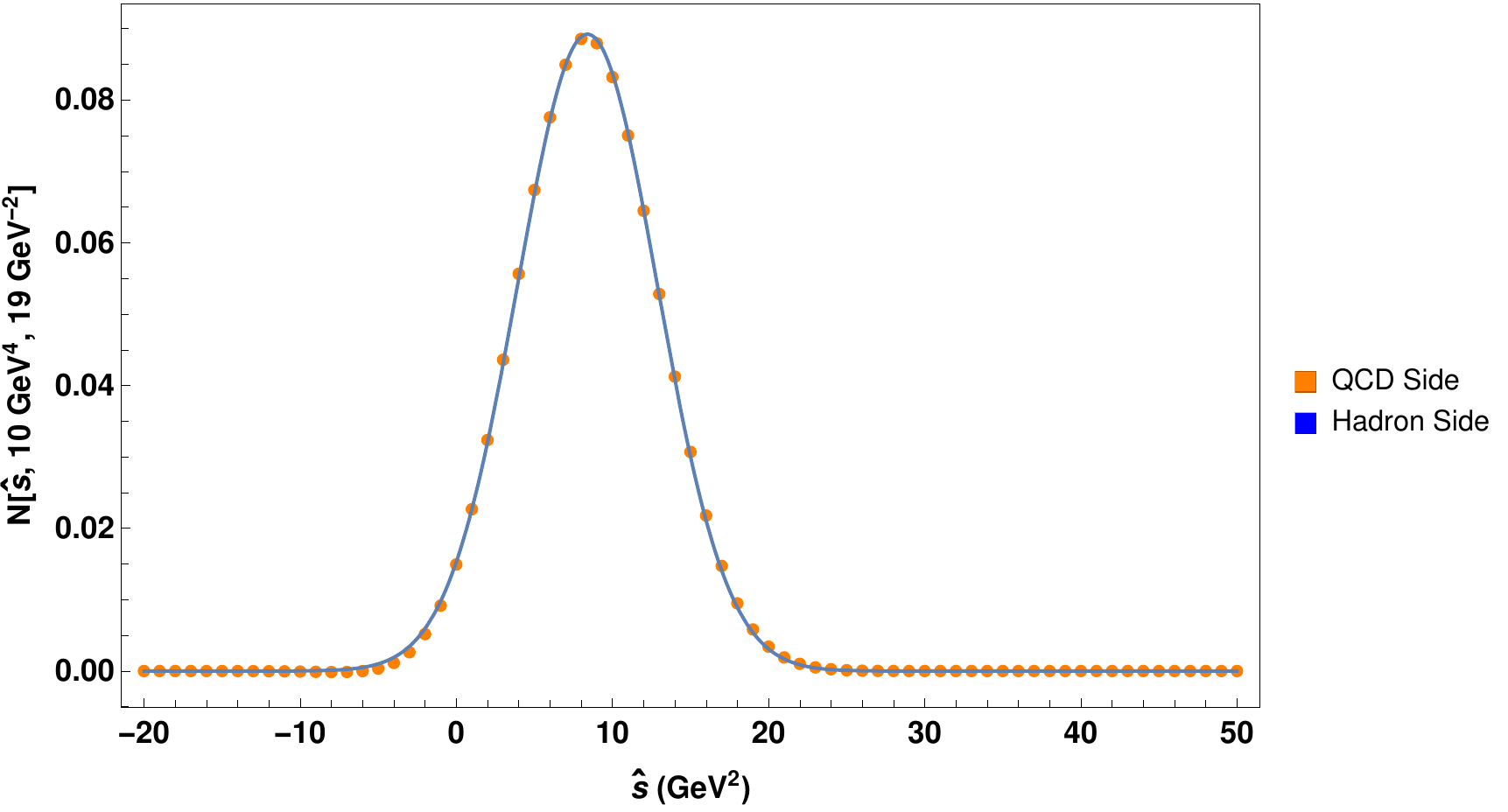}
    \caption{Double-narrow resonance model 
    $N^\text{had}\left( \hat{s},\,\tau \right)$ (solid blue curve) and compared to QCD prediction $N^\qcd(\hat{s},\,\tau,\,s_0)$ (orange points) for $\tau=10\,\gev^4$.  Central values of the QCD condensates and 
    the corresponding fitted parameters have been used.
    }
    \label{fig:model}
\end{figure}

We choose the width of the Gaussian kernel to be $ \tau = 10 \, \gev^4,$ in line with our previous GSRs analysis of light hybrids \cite{Ho:2018cat}. Since this resolution is much larger than the experimental width of the $Y(2175)$ (\textit{i.e.,} $\sqrt\tau\gg m_1\Gamma $), the narrow width model is an excellent approximation for the $Y(2175)$.  For the undetermined resonance $m_2$, we assume that it is similarly narrow compared to the Gaussian kernel resolution; this assumption is revisited in the results of our analysis presented below.
To determine the remaining unknown quantities $\{m_2, r, s_0\}$ in our model we seek the best fit of the the $\hat s$ dependence of the QCD prediction and hadronic model  by minimizing the $\chi^2$,
\begin{equation}
\chi^2\left(r,m_2,s_0 \right)=\sum_{\hat s_{min}}^{\hat s_{max}}
\left[ N^\text{had}\left( \hat{s},\,\tau \right)- N^\text{QCD}\left( \hat{s},\,\tau\,s_0 \right) \right]^2\,,
\label{chi2}
\end{equation}
where we use 161 equally spaced $\hat s $ points with 
$\hat s_{\text{min}}=-10\,\gev^4$ and  $\hat s_{\text{max}}=30\,\gev^4$.  This region safely encloses the resonances resulting from our analysis as outlined below.
Note that the minimization is constrained by the physical condition $0\le r\le 1$ in \eqref{rDefn}.  The resulting prediction of the resonance parameters and continuum onset is 
\begin{gather}\label{fitResults}
    s_0^{\mathrm{opt}} = 9.7\pm1.0\,\gev^2\\
    m_2 = m_\mathrm{fit} = 2.90\pm0.16 \,\gev\\
    r \le 0.033\,\label{fitResults_r}.
\end{gather}
The uncertainties in \eqref{fitResults}--\eqref{fitResults_r} are obtained by varying the values of the QCD input parameters, and calculating the deviation from the central values in quadrature. Errors are dominated by the variation in $\langle \alpha G^2 \rangle$.  An upper bound on $r$ is provided because of the  $r\ge 0$ constraint.  Figure \ref{fig:model} shows that the  fit between the QCD prediction and hadronic model is excellent; there is no evidence of any deviations that would suggest a need to refine the model (\textit{e.g.,} inclusion of a numerically-large width $\sqrt\tau\sim m_2\Gamma$ for $m_2$).  Figure \ref{fig:model} also shows that the fitted region $ -10 \, \gev^2 <\hat s <30\, \gev^2$ encloses the regions where the NGSRs are numerically significant.  As a further validation of our results, we note that our mass prediction for $m_2$ is consistent with previous LSRs analyses~\cite{Govaerts:1985fx}. 

\begin{figure}
    \centering
    \includegraphics[scale=0.5]{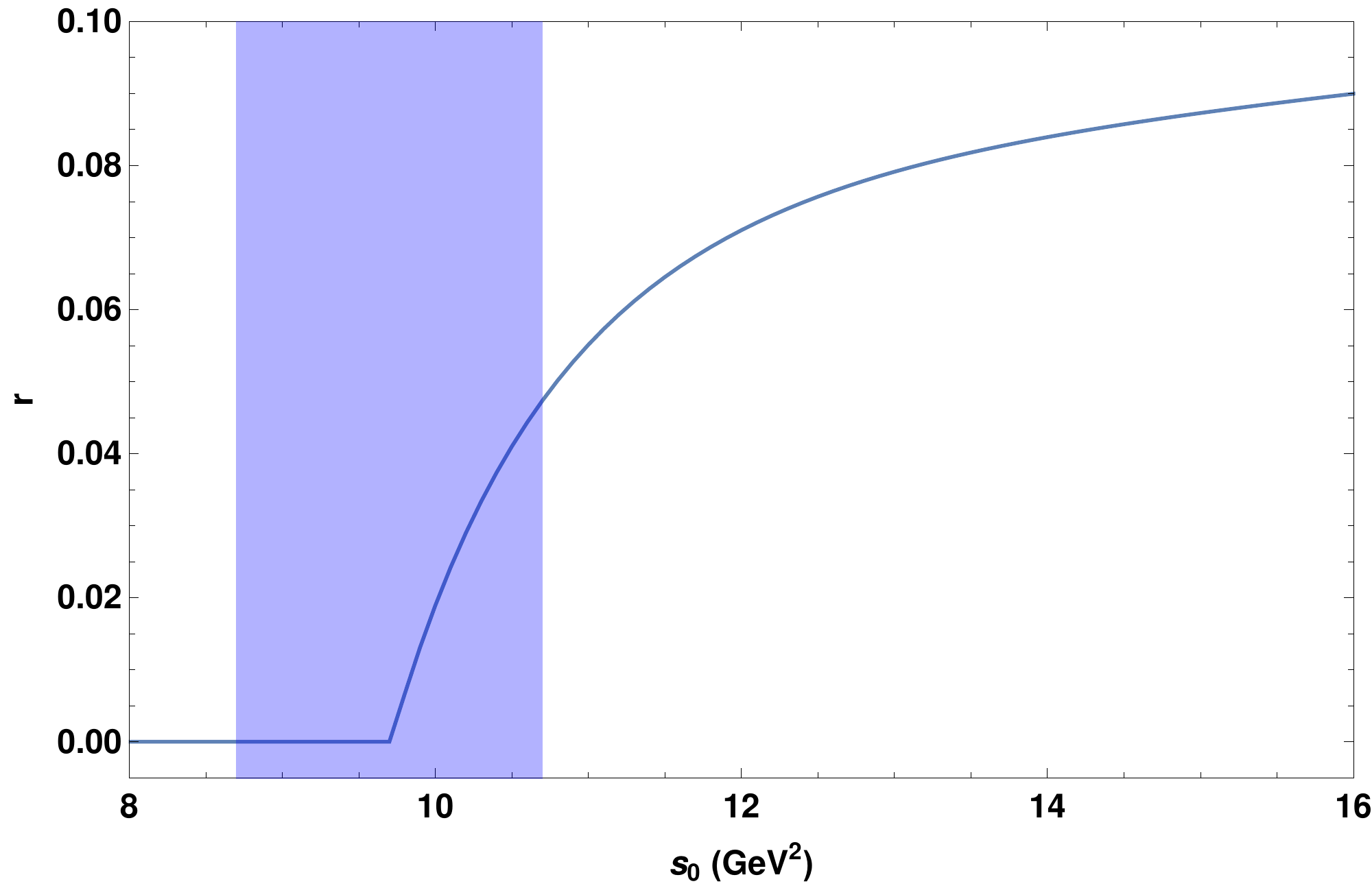}
    \caption{Predicted coupling $r$ to $Y(2175)$ state 
    as a function of the continuum onset $s_0$. The vertical band highlights the optimized value of continuum onset $s_0^\mathrm{opt}$ with corresponding 
    error~(\ref{fitResults}). The physical constraint $r>0$ has been imposed in the analysis.}
    \label{fig:r(s0)bands}
\end{figure}

\begin{figure}
    \centering
    \includegraphics[scale=0.5]{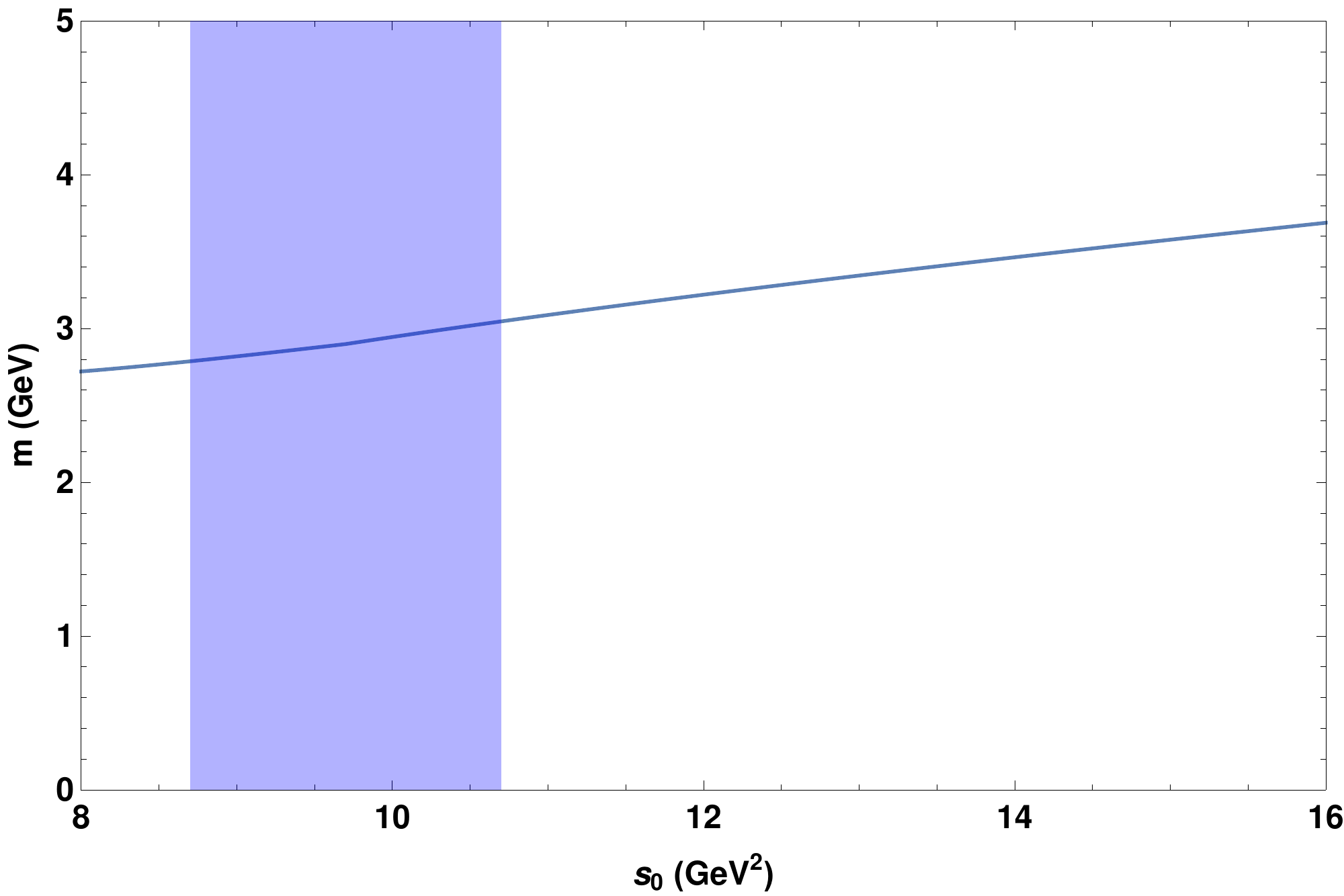}
    \caption{Predicted vector strangeonium hybrid mass  $m_2$ as a function of the continuum onset $s_0$. The vertical band highlights the optimized value of continuum onset $s_0^\mathrm{opt}$ with corresponding error~(\ref{fitResults}).}
    \label{fig:m(s0)bands}
\end{figure}

The key aspect of our results \eqref{fitResults}--(\ref{fitResults_r})
is the small relative resonance  strength  $r\leq3.3\%$ of the $Y(2175)$ compared to $m_2$, which seems to preclude a predominant hybrid component of the $Y(2175)$.  We can obtain a more conservative bound on $r$ by calculating the  $s_0$ dependence of $r$ (\textit{i.e.,} choosing $s_0$ and only fitting $r$ and $m_2$) and then considering the variation of $r$ within the  region of uncertainty  
in $s_0$ from~(\ref{fitResults}).
The result of this analysis leads to the bound $r\leq5\%$
as shown in Figure~\ref{fig:r(s0)bands}.  A similar analysis for $m_2$ is shown in Figure~\ref{fig:m(s0)bands}.

\section{Discussion}\label{V}
In summary, we have used QCD GSRs to study the strangeonium hybrid interpretation of the $Y(2175)$.
Compared to a previous LSRs analysis of vector strangeonium hybrids~\cite{Govaerts:1985fx}, our calculation includes 5d and 6d condensate contributions, strange quark mass corrections to perturbation theory, and updated QCD parameter values.
Furthermore, the advantage of the GSRs approach over the LSRs approach is its comparable 
sensitivity to multiple states in a hadronic spectral function.  
This allowed us to explore the relative coupling to the hybrid current~\eqref{CurHyb} of the $Y(2175)$ and an additional unknown resonance. 
We found excellent agreement between the QCD prediction and hadronic model, 
and determined an upper bound $r\leq5\%$ for the relative coupling strength of the $Y(2175)$.  
In other words, we found no evidence for a significant strangeonium hybrid component of the $Y(2175)$.

Recently, a structure of mass $(2239\pm13.3)$~MeV and width $(139.8\pm24.0)$~MeV
(where we have combined statistical and systematic uncertainties) 
was observed in $e^{+}\,e^{-}\,\rightarrow\,K^{+}\,K^{-}$ with the BES III detector~\cite{Ablikim:2018iyx}.
If the structure can be identified with the $Y(2175)$, then the observed $KK$ decay mode
would disfavour the $3\,{}^{3}S_{1}$ strangeonium meson, strangeonium hybrid, and $ss\bar{s}\bar{s}$ 
tetraquark interpretations.
On the other hand, if the structure can not be identified with the $Y(2175)$, then the lack of observed
$KK$ decay mode would disfavour the $2\,{}^{3}D_1$ strangeonium meson and $\Lambda\bar{\Lambda}$ interpretations. 
Clearly, further experimental and theoretical studies are needed.

\section*{Acknowledgments}
We are grateful for financial support from the Natural Sciences and 
Engineering Research Council of Canada (NSERC), and the Chinese National Youth Thousand Talents Program.
\clearpage
\bibliography{main}

\end{document}